# AN EXPLICIT SOLUTION FOR AN OPTIMAL STOPPING/OPTIMAL CONTROL PROBLEM WHICH MODELS AN ASSET SALE[1]

By Vicky Henderson and David Hobson

*Warwick Business School and University of Warwick*

In this article we study an optimal stopping/optimal control problem which models the decision facing a risk-averse agent over when to sell an asset. The market is incomplete so that the asset exposure cannot be hedged. In addition to the decision over when to sell, the agent has to choose a control strategy which corresponds to a feasible wealth process.

We formulate this problem as one involving the choice of a stopping time and a martingale. We conjecture the form of the solution and verify that the candidate solution is equal to the value function.

The interesting features of the solution are that it is available in a very explicit form, that for some parameter values the optimal strategy is more sophisticated than might originally be expected, and that although the setup is based on continuous diffusions, the optimal martingale may involve a jump process.

One interpretation of the solution is that it is optimal for the risk-averse agent to gamble.

**1. Introduction.** The aim of this article is to study a mixed optimal stopping/optimal control problem which arises in a natural way in finance as a mixed investment/sale problem in an incomplete market.

The mathematical problem and its solution have several interesting features. First, we can find an explicit solution and this is especially pleasing since the problem is multidimensional, although it does possess certain natural scalings. Second, the form of the solution is unexpected, and turns out to be more complicated than naive intuition might expect. In a sense to be made precise later, even though the objective function is concave, the

Received May 2007; revised August 2007.
[1]Supported by an EPSRC Advanced Fellowship.

choice over stopping times induces an unexpected convexity. Third, the optimal strategy has novel mathematical features, and although the setup is based on continuous processes, the optimal control involves a combination of jumps and local times on rays.

There is a long history of problems from finance, such as those concerning portfolio optimization (Merton [9]) and in American options (Samuelson/McKean [12]) being converted into problems in stochastic control and optimal stopping. Examples of mixed stopping/control problems include Henderson [3], Karatzas and Kou [5], Karatzas and Sudderth [7], Karatzas and Wang [8] and Karatzas and Ocone [6].

The situation that motivates our study is that of a risk-averse[2] agent who has an asset to sell at a time of her choosing, and who wishes to choose the sale time so as to maximize the expected utility of total wealth at that moment. Here total wealth includes both the revenue from the sale, and the value of any other assets. This is an example of a problem from "Real Options" (see, e.g., Dixit and Pindyck [1]) where an asset is described as real in order to distinguish it from a financial asset which may be dynamically hedged on a market. In the language of financial economics, the asset is not redundant,[3] and the market is incomplete. Although our problem can be stated as an abstract stochastic control problem, we will often use the language of financial mathematics, such as describing the objective function as a utility function, and the controlled process as a wealth process. Our intuition for the solution will partly come from the application to finance.

We consider two versions of the problem. In the simpler version, the agent has a simple optimal stopping problem of when to sell the real asset, and other wealth is constant. (We work with the bank account as numeraire, and imagine in that case that the agent has no outside investment opportunities.) In the extended version, the agent has the possibility to invest her other wealth. If she could choose investments which displayed correlation with the real asset, then she could reduce her risk exposure, by selling the financial asset if necessary. Similarly, if she could choose investments with nonzero expectation, then she could increase her expected returns. In either case we would expect her to be able to increase her expected utility.

We rule out both of these possibilities by insisting that the set of available investments are uncorrelated with the real asset, and that they are fair. (In terms of the mathematical statement of the problem, this means that the set of available controls are martingales which have zero covariation with respect to the price process of the real asset.) At first sight it appears that there is no way that an agent can make use of these extra opportunities. However, we find that this is not the case, and that the presence of the timing option to sell the real asset means that the risk-averse agent becomes risk-seeking. Effectively the agent chooses to gamble. Thus our model provides a rational explanation for gambling, albeit in a specialized setting, without recourse to nonconcave utilities, inaccurate assessments of probabilities, or other irrationalities.

---

[2] For our purposes (see also Müller and Stoyan [10]) a risk-averse agent is one who, offered the choice between the certain yield $\mathbb{E}[Z]$ and the risky prospect $Z$, both payments to be made at a future time $T$, prefers the certain payout, whatever the distribution of $Z$.

[3] An asset is redundant if its payoffs can be replicated through other securities.

The remainder of the paper is structured as follows. In the next section we give a mathematical formulation of the problem and state the main result. The problem with no investment opportunities is the subject of Section 3, and the situation where additional fair investments or gambles are available is contained in Section 4. In both of these sections we consider an agent with a utility with constant relative risk aversion coefficient less than 1. The results are extended to the case $R > 1$ in Section 5. This involves considerable extra work as an easy proof that a local martingale is a supermartingale which is valid for $R < 1$ is no longer valid in this case. Section 6 contains the logarithmic utility case together with concluding remarks.

## 2. Utility maximization with discretionary stopping.

2.1. *Preliminaries.* The generic problem (in the next subsection we will be both more precise and more explicit) is to find

$$(1) \qquad V_* = \sup_{\tau, X \in \mathcal{X}} \mathbb{E}[U(\tau, X_\tau, Y_\tau)]$$

where $\tau$ is a stopping time, $X_t$ is a stochastic control chosen from a space $\mathcal{X}$ of feasible strategies, and $Y$ is an exogenous Markov process. In the terminology of Karatzas and Wang [8] the stopping time $\tau$ is discretionary in the sense that it is chosen by the agent.

The problem (1) should be compared with the problem in Karatzas and Wang [8] [see (5.1) and (5.3)]. In some ways (1) is simpler, not least because it focuses completely on utility of terminal wealth, whereas the formulation in Karatzas and Wang includes utility from consumption. (However, in terms of the general setup of this subsection, it is clear that we too could include a consumption term.) A further difference is that Karatzas and Wang place an explicit interpretation on $X$ as the gains from trade from an investment strategy in a multi-asset, frictionless market. Again, this difference is largely cosmetic, and although we are not similarly explicit, this interpretation is also the one we have in mind. The main motivation for the more general definition of $\mathcal{X}$ is so that optimal strategies will exist.

Instead, there are two fundamental differences between the problem in (1) and the problem in Karatzas and Wang [8]. First, these latter authors assume that $U$ has no $Y$ dependence and takes the particular form $U(t, x, y) \equiv$ $e^{-\rho t} U(x)$. Second, they work in a complete market in which the traded financial assets span the uncertainty set in the model. In contrast, in our setup the auxiliary process $Y$ (representing the price of the asset to be sold) enters into the optimization problem, but is not part of the financial market, so that the model is not complete. Indeed, the formulation (1) can itself be seen as a special case of a more general problem in which $U$ is random.

2.2. *Problem definition.* The idea is that $X$ represents the wealth process of an agent, and that $Y$ represents the price process of a real asset. The agent owns a single unit of the real asset. At the moment $\tau$ of the agent's choosing at which the real asset is sold, the wealth of the agent increases from $X_{\tau-}$ to $X_\tau + Y_\tau$. The objective of the agent is to maximize the expected (increasing, concave) utility of wealth, evaluated at $\tau$.

Let $Y$ be modeled by an exponential Brownian motion with dynamics

$$(2) \qquad dY_t = Y_t(\sigma\, dW_t + \mu\, dt), \qquad Y_0 = y > 0,$$

which is specified exogenously. It is convenient to set $\mu = \gamma \sigma^2/2$.

The fundamental problem is to select a stopping time $\tau$ and a stochastic process $X_t$ from the set $\mathcal{X}$ where

$$(3) \qquad \begin{aligned} \mathcal{X} = \{X_t : \mathbb{E}[X_0] = x; X_u + Y_u \geq 0,\ u \leq \tau; \\ X \text{ is a càdlàg martingale such that } [X,Y]_t \equiv 0\} \end{aligned}$$

so as to solve (for $y > 0$ and $x > -y$)

$$(4) \qquad V_*^g \equiv V_*^g(x,y) = \sup_{\tau, X \in \mathcal{X}} \mathbb{E}[U_R(X_\tau + Y_\tau)|X_0 = x, Y_0 = y],$$

where the objective function $U$ is given by

$$(5) \qquad U(u) = U_R(u) = \frac{u^{1-R} - 1}{1 - R}, \qquad R \in (0,\infty),\ R \neq 1.$$

In particular we want to compare the solution to (4) with

$$(6) \qquad V_*^n \equiv V_*^n(x,y) = \sup_\tau \mathbb{E}[U_R(x + Y_\tau)|Y_0 = y],$$

where again the domain of $V_*^n$ is $(y > 0, x + y > 0)$. If $x \geq 0$ in (6), then $\tau$ is unrestricted; otherwise for $x < 0$ we are only interested in the case $x + y > 0$ and we insist that $\tau \leq \inf\{u : Y_u = -x\}$.

Note that we do not work on a given probability space, but rather we are allowed to design a model $(\Omega, \mathcal{F}, \mathbb{P})$ with associated filtration $\mathbb{F}_t$, provided we do this in a way such that $W$ is an $\mathbb{F}_t$-Brownian motion. (In the same spirit Karatzas and Ocone [6] consider the probabilistic model as part of the solution, and call such a model policy supporting.)

The superscript $g$ on the value function denotes the fact that the (wealth) process $X$ is a martingale, or equivalently that the process $X$ is the wealth resulting from a series of fair gambles. The superscript $n$ is an abbreviation for the no-gambling case. Furthermore, the slightly nonstandard form of the utility function is chosen so that in the limit $R \to 1$ we recover logarithmic utility: recall that $\lim_{R \to 1}(x^{1-R} - 1)/(1-R) = \ln x$. As a result the case of logarithmic utility $U(x) = \ln x$ can be recovered immediately in the limit $R \to 1$.

The solution to (2) is given by $Y_t = y e^{\sigma W_t + (\gamma - 1)\sigma^2 t/2}$ where $\gamma = 2\mu/\sigma^2$. The interesting case is when $0 < \gamma < 1$. In this case $Y$, which is our model for the real asset, is increasing in expectation, but does not increase to infinity almost surely. The fact that $Y$ has positive drift gives an incentive to hold onto the real asset, which is counterbalanced by the risk of price fluctuations resulting from a delayed sale, to give a nondegenerate problem. Further, in the case $\gamma < 1$ we have that $Y$ tends to zero almost surely.

Note that in both (4) and (6) we do not want to insist that stopping time $\tau$ is finite [and indeed, for $0 < \gamma < 1$ it will turn out that $0 < \mathbb{P}(\tau < \infty) < 1$ for the optimal stopping rule]. However, we do need to be sure that $\lim(X_{t \wedge \tau} + Y_{t \wedge \tau})$ and $\lim(x + Y_{t \wedge \tau})$ exist as $t$ increases to infinity. The existence of the latter limit is guaranteed for $\gamma < 1$ since $Y_t \to 0$ almost surely as $t \to \infty$. Further, the condition $X_{t \wedge \tau} + Y_{t \wedge \tau} \geq 0$ provides a bound on $X$ from below, and as we shall see this is sufficient to give a limit for $(X_{t \wedge \tau} + Y_{t \wedge \tau})$.

The definition of the set $\mathcal{X}$ of admissible wealth processes reflects several features. The fact that the quadratic variation $[X, Y] = 0$ is a formalization of the idea that there are no hedging instruments for $Y$. The martingale assumption reflects the fact that the wealth process $X$ is the gains process from a series of fair gambles. (In the financial setting it might be appropriate to replace the martingale condition with a condition that $X$ is a supermartingale. It should be clear that this does not change the results.) The condition $X_u + Y_u \geq 0$ for $u \leq \tau$ is a natural restriction and rules out doubling strategies and gambling for resurrection. Finally, the càdlàg assumption is a regularity condition to assume that certain quantities are well defined.

Note that many of these assumptions can be relaxed but at the cost of losing the explicit solution, and of losing the context within which our result (see Corollary 3 below) is so counterintuitive. For example (see Evans, Henderson and Hobson [2]) if we suppose that $X$ is the gains from trade from an investment strategy in a traded asset $P$, then we may have that $P$ (and thence $X$) is correlated with $Y$, or that $P$ (and thence $X$) has nonzero drift.

2.3. *Main results.* In this section we state the main result in the case $R < 1$. The proof of this result will follow from the analysis in the next two sections.

Recall that we are interested in the solutions to (4) and (6), and especially when these two solutions are different. It turns out that there is a transition in the form of the solution at a particular value of $\gamma$, namely $\gamma_-$.

DEFINITION 1. Let $\Gamma_R(\gamma)$ be given by
$$\Gamma_R(\gamma) = (R - \gamma)^R(R + 1 - \gamma) - (2R - \gamma)^R(1 - \gamma),$$
and let $\gamma_- \equiv \gamma_-(R)$ be the unique solution in $(0, R \wedge 1)$ of $\Gamma_R(\gamma) = 0$.

For two value functions $V^{(1)}, V^{(2)}$ defined on $(y > 0, x + y \geq 0)$ we write $V^{(1)} \equiv V^{(2)}$ if $V^{(1)}(x,y) = V^{(2)}(x,y)$ for all pairs $(x,y)$ and $V^{(1)} < V^{(2)}$ if $V^{(1)}(x,y) \leq V^{(2)}(x,y)$ for all pairs $(x,y)$ with strict inequality for some pair.

THEOREM 2. *Suppose $R < 1$. For $\gamma \leq \gamma_-(R)$ (and for $\gamma > R$) we have that $V_*^n \equiv V_*^g$. Conversely, for $\gamma_- < \gamma \leq R$ we have that $V_*^n < V_*^g$.*

The results in the case $R \geq 1$ are broadly similar; see, for example, Theorem 14. However, the fact that the objective function is not bounded below when $R \geq 1$ introduces some serious complications. For this reason we defer consideration of this case until we have studied the case $R < 1$ in full detail.

In terms of the financial asset sale problem, Theorem 2 has the following corollary:

COROLLARY 3. *Suppose $R < 1$. For $\gamma \leq \gamma_-(R)$ the solution to the optimal selling problem is the same whether or not the agent has access to fair gambles. For $\gamma_- < \gamma \leq R$ the risk-averse agent can improve her utility by undertaking fair gambles.*

The main result of Corollary 3, namely that the agent can benefit from the opportunity to make alternative investments, would not be a surprise if either the investments facilitated hedging, or if the investments were beneficial in their own right. However, by insisting that these additional investments have zero return, and that they are uncorrelated with the real asset, we have ruled out both of these motives for outside investment. If the real asset value $Y$ was constant, or if $\tau$ was fixed and predetermined, then there would never be a benefit to be obtained from nonconstant $x$, and the agent would never gamble.

Instead, there are three key elements in our model which are necessary for the main conclusion that gambling can be beneficial. These are that the market is incomplete, so that exposure to fluctuations in $Y$ cannot be fully hedged, that the real asset is indivisible, and that the asset sale is irreversible. However, it remains a surprise that these features are sufficient to induce a risk-averse agent to gamble.

**3. Calculation of solution with constant wealth.** We begin by deriving the solution to (6) in the case $0 < R < 1$. Rather than writing down a variational problem (HJB equation) we postulate that the optimal stopping rule belongs to a natural candidate family. We calculate the value function for the optimal element of this family, and then use a verification argument to show that we have solved the original problem. In fact the class of stopping rules we will consider are the first times that $Y$ exceeds some large level.

Given the time-homogeneity of the problem, it is natural that the optimal stopping region should be independent of time. Further, increased $Y$ is associated with greater risks so a natural candidate for the optimal stopping time is the first time that $Y$ exceeds a given level. To this end, for $0 < w \leq x/y$ define
$$\tau_w = \inf\{u \geq 0 : Y_u \geq x/w\}.$$

Suppose $\gamma \leq 0$. Then, for all $\tau$, $\mathbb{E}[U(x + Y_\tau)] \leq U(x + \mathbb{E}[Y_\tau]) \leq U(x + y) = \mathbb{E}[U(x + Y_0)]$. Thus $\tau = 0$ is optimal. In financial terms the asset $Y$ is depreciating and since there is also a risk inherent in waiting to sell the asset and $U$ is concave, it is optimal to choose $\tau$ as small as possible.

Now suppose $\gamma \geq 1$. Suppose $x > 0$. Then $\tau_w < \infty$ for arbitrarily small $w$ and $V_*^n(x,y) \geq U(x(1+1/w))$. It follows that $V_*^n(x,y) = \infty$. To deal with the case $x < 0$, for $0 < v < |x|/y$ define
$$\tilde{\tau}_v = \inf\{u \geq 0 : Y_u = |x|/v\},$$
and set $\tilde{\tau}_1 = \inf\{u \geq 0 : Y_u = |x|\}$. Then, for $v < |x|/y$,

(7)
$$\mathbb{E}[U(x + Y_{\tilde{\tau}_1 \wedge \tilde{\tau}_v})]$$
$$= -\frac{1}{1-R} + \frac{1}{1-R}(|x|^{1-R}(-1 + 1/v)^{1-R})\mathbb{P}(\tilde{\tau}_v < \tilde{\tau}_1).$$

We can use the scale function of $Y$ to calculate (for $\gamma > 1$), $\mathbb{P}(\tilde{\tau}_v < \tilde{\tau}_1) = (|x|^{1-\gamma} - y^{1-\gamma})/(|x|^{1-\gamma}(1 - v^{\gamma-1}))$, and then $\mathbb{E}[U(x + Y_{\tilde{\tau}_1 \wedge \tilde{\tau}_v})]$ tends to infinity as $v$ tends to zero. [For $\gamma = 1$ we have $\mathbb{P}(\tilde{\tau}_v < \tilde{\tau}_1) = (\ln y - \ln|x|)/(-\ln v)$ with the same conclusion.] Hence $V_*^n(x,y) = \infty$ for $\gamma \geq 1$.

The case $R < \gamma < 1$ is also degenerate in a similar fashion. Define $F(w) = \mathbb{E}[U(x+Y_{\tau_w})]$, and recall that for $\gamma < 1$, $Y_t \to 0$. Then, for $x > 0$ and $y \leq x/w$,

(8)
$$F(w) = \frac{(x^{1-R} - 1)}{1-R}\mathbb{P}(\tau_w = \infty) + \frac{x^{1-R}(1+1/w)^{1-R} - 1}{1-R}\mathbb{P}(\tau_w < \infty)$$
$$= \frac{x^{1-R}[1 + \{(1+1/w)^{1-R} - 1\}(wy/x)^{1-\gamma}] - 1}{1-R}.$$

On differentiating we have
$$F'(w) = \left(\frac{y}{x}\right)^{1-\gamma} x^{1-R} w^{-\gamma}\left[(1-\gamma)\frac{(1+1/w)^{1-R} - 1}{1-R} - \frac{(1+1/w)^{-R}}{w}\right]$$

and since, for $w > 0$

$$\frac{(1+1/w)^{-R}}{w} > (1+1/w)^{1-R} - 1,$$

it follows that for $R \leq \gamma < 1$, $F$ is a decreasing function. Then $V_*^n(x,y) \geq \lim_{w \downarrow 0} F(w) = \infty$. If $x < 0$, the argument of the previous paragraph still applies and $V_*^n(x,y) = \infty$. The case $x = 0$ is also easily dealt with.

If $\gamma = R < 1$ and $x > 0$, then $V_*^n(x,y) \geq \lim_{w \downarrow 0} F(w) = (x^{1-R} + y^{1-R} - 1)/(1-R)$. Conversely, it follows from the twin facts that $(x+y)^{1-R} \leq x^{1-R} + y^{1-R}$ and that $Y^{1-R}$ is a nonnegative martingale, that $\mathbb{E}[U_R(x+Y_\tau)] \leq \mathbb{E}[(x^{1-R} + Y_\tau^{1-R} - 1)/(1-R)] \leq (x^{1-R} + y^{1-R} - 1)/(1-R)$. Hence $V_*^n(x,y) = (x^{1-R} + y^{1-R} - 1)/(1-R)$. Further, if $x \leq 0$, then it is easy to check using Itô's formula that $(x+Y_t)^{1-R}$ is a positive supermartingale, and hence $V_*^n(x,y) = ((x+y)^{1-R} - 1)/(1-R)$.

It remains to consider the nondegenerate case, which is covered in the following results.

DEFINITION 4. For $0 < w < \infty$ define $\Lambda(w)$ by

$$\Lambda(w) = (1-\gamma)\frac{(1+1/w)^{1-R} - 1}{1-R} - \frac{(1+1/w)^{-R}}{w}. \tag{9}$$

LEMMA 5. For $0 < \gamma < R < 1$ there is a unique solution $w^*$ to $\Lambda(w) = 0$, and $w^* < (R-\gamma)/\gamma$.

PROOF. On differentiation we see that $\Lambda$ has a unique turning point at $w = (R-\gamma)/\gamma$, and that this turning point is a minimum. Further, $\lim_{w \to 0} \Lambda(w) > 0 = \lim_{w \to \infty} \Lambda(w)$. It then follows that $(R-\gamma)/\gamma > w^*$ where $w^*$ is the unique positive solution to $\Lambda(w) = 0$. □

For $w > 0$ define $V^w$ via: for $y \geq x/w$,

$$V^w(x,y) = \frac{(x+y)^{1-R} - 1}{1-R}; \tag{10}$$

and for $y < x/w$,

$$V^w(x,y) = \frac{x^{1-R} - 1}{1-R} + \left(\frac{yw}{x}\right)^{1-\gamma} x^{1-R}\frac{(1+1/w)^{1-R} - 1}{1-R}. \tag{11}$$

Note that for $y < x/w$, $V^w(x,y) = F(w)$.

PROPOSITION 6. Suppose $R < 1$.
For all $\gamma \leq 0$, $V_*^n(x,y) = ((x+y)^{1-R} - 1)/(1-R)$.
For all $\gamma > R$, $V_*^n(x,y) = \infty$.

For $\gamma = R$, $V_*^n(x,y) = (x^{1-R} + y^{1-R} - 1)/(1-R)$ for $x \geq 0$ and $V_*^n(x,y) = ((x+y)^{1-R} - 1)/(1-R)$ for $x < 0$.

In the nondegenerate cases $0 < \gamma < R$, $V_*^n \equiv V^{w^*}$, where $w^*$ is the unique positive solution to $\Lambda(w) = 0$.

PROOF. We cover the case $0 < \gamma < R$, the other cases having been covered in the discussion before the statement of the proposition. We need to show that $V^{w^*} \equiv V_*^n$, where $V_*^n$ is the solution to the problem in (6).

*Lower bound.* If $x \leq w^* y$, then $V^n(x,y) \geq U(x+y) = V^{w^*}(x,y)$. Otherwise $x$ is certainly positive, and for $w < x/y$ and for the stopping rule $\tau_w$ we have $\mathbb{E}[U(x + Y_{\tau_w})] = F(w)$ where $F$ is as given in (8). For $0 < \gamma < R$ the equation $F'(w) = 0$ has a unique solution $w^*$ in $(0, \infty)$, and it is easy to see that this solution corresponds to a maximum of $F$. Then, for $x \geq w^* y$, $V_*^n(x,y) \geq F(w^*) = V^{w^*}(x,y)$.

*Upper bound.* It is an exercise in one-dimensional calculus to show that $V^{w^*}$ satisfies (a subscript $y$ denotes partial differentiation)

$$(12) \qquad V^{w^*}(x,y) \geq U_R(x+y),$$

$$(13) \qquad \mu y V_y^{w^*} + \tfrac{1}{2}\sigma^2 y^2 V_{yy}^{w^*} \leq 0,$$

with equality in (12) for $y \geq x/w^*$, and equality in (13) for $y < x/w^*$. In particular, for $y > x/w^*$

$$\gamma y V_y^{w^*} + y^2 V_{yy}^{w^*} = (x+y)^{-(R+1)} y^2 \left\{ \gamma \frac{x}{y} - (R - \gamma) \right\}$$

and this expression is negative, since by Lemma 5, $w^* < (R-\gamma)/\gamma$. Further, using the definition of $w^*$, $V_y^{w^*}$ is continuous at $y = x/w^*$ and equal to $x^{-R}(1 + 1/w^*)^{-R}$.

By Itô's formula, for any pair of initial values $(x,y)$ and for any stopping time $\tau$,

$$V^{w^*}(x, Y_{t \wedge \tau}) = V^{w^*}(x,y) + \int_0^{t \wedge \tau} (\mu Y_s V_y^{w^*}(x, Y_s) + \tfrac{1}{2}\sigma^2 Y_s^2 V_{yy}^{w^*}(x, Y_s)) \, dt$$

$$+ \int_0^{t \wedge \tau} \sigma Y_t V_y^{w^*}(x, Y_s) \, dW_s$$

$$\leq V^{w^*}(x,y) + \int_0^{t \wedge \tau} \sigma Y_s V_y^{w^*}(x, Y_s) \, dW_s.$$

The right-hand side of this expression is a local martingale which is bounded below [by $V^{w^*}(0,0) = -1/(1-R)$] and hence is a supermartingale. Then, using (12),

$$\mathbb{E}^y[U_R(x + Y_{t \wedge \tau})] \leq \mathbb{E}^y[V^{w^*}(x, Y_{t \wedge \tau})]$$

$$\leq V^{w^*}(x,y) + \mathbb{E}^y\left[\int_0^{t\wedge\tau} \sigma Y_s V_y^{w^*}(x,Y_s)\,dW_s\right]$$

$$\leq V^{w^*}(x,y).$$

Letting $t \uparrow \infty$ and using Fatou we conclude $V_*^n(x,y) \leq V^{w^*}(x,y)$. □

**4. Calculation of the solution with gambling.** The strategy of this section is similar to that in the previous section. We exhibit a parametric family of combined stopping rules and admissible martingales $X_t$. For each element in the family we calculate the associated value function, and then we optimize over the parameter values. This gives a lower bound on $V_*^g$. Finally we show that this lower bound is also an upper bound.

Recall that we are assuming $R < 1$. In order to rule out degenerate solutions we assume that $\gamma < 1$.

4.1. *Definition of a family of candidate strategies.* Fix $-1 < \xi < \eta$ with $\eta > 0$. The aim is to specify a stopping rule $\tau = \tau^{\eta,\xi}$ and a martingale $X = X^{\eta,\xi}$.

Suppose first that $x \leq \xi y$. In this case we take $\tau = 0$, and $X_0 = x$. Now suppose $\xi y < x < \eta y$. Let $X_0$ be the random variable such that $X_0 = \eta y$ with probability $(x - \xi y)/((\eta - \xi)y)$ and $X_0 = \xi y$ otherwise. Then $\mathbb{E}[X_0] = x$. On $X_0 = \xi y$, take $\tau = 0$. Finally, if $x \geq \eta y$, set $X_0 = x$. In all cases we have that the sets $\tau > 0$ and $X_0 \geq \eta y$ are identical.

Hence we may suppose that $X_0 \geq \eta y$. Define $L_t = (\max_{s \leq t} Y_s - X_0/\eta)^+$. [We use the label $L$, since by Skorokhod's lemma, $L$ is proportional to the local time the process $(X_t, Y_t)$ spends on the ray $X_s = \eta Y_s$, where $X$ is defined below.] Note that $L$ is an increasing process and define

$$A_t = \frac{\eta}{(\eta-\xi)} \int_0^t \frac{dL_u}{Y_u}.$$

Let $N$ be a standard Poisson process, independent of $Y$, and define $X_t$ via

$$(14) \quad X_t = X_0 + \int_0^t I_{\{N_{A_{s-}}=0\}} \eta\,dL_s - \int_0^t I_{\{N_{A_{s-}}=0\}} Y_s(\eta-\xi)\,dN_{A_s}.$$

It follows that $[X,Y]_t \equiv 0$. Associated with $X_t$ will be the stopping rule

$$(15) \quad \tau^{\eta,\xi} = \sup\{u : N_{A_u} = 0\},$$

so that if $X$ ever jumps, then the real asset is sold. For this reason we concentrate on the process while $N_{A_{t-}} = 0$.

Provided $N_{A_t} = 0$ we have that $X_t = X_0 + \eta L_t$. It follows that $X$ only increases when $L$ increases (which can only occur when $Y$ is at a maximum), and at those times $L_t = Y_t - X_0/\eta$, so that $X_t = \eta Y_t$. Then, taking the derivative form of (14), and substituting for $dL_t$,

$$dX_t = (\eta\, dL_t - Y_t(\eta - \xi)\, dN_{A_t}) I_{\{N_{A_{t-}} = 0\}}$$
$$= -Y_t(\eta - \xi)(dN_{A_t} - dA_t) I_{\{N_{A_{t-}} = 0\}}.$$

It follows immediately that $X$ is a local martingale, and the true martingale property follows since $X$ is bounded above and below by constant multiples of the maximum of the exponential Brownian motion $Y_t$.

Furthermore, it is also the case that the value of $X$ can decrease only when $X_{t-} = \eta Y_t$. To see this, note that $N_{A_t}$ can only jump at a point of increase of $A_t$, and this can only happen when $L$ is increasing, so that as before (but now using the right-continuity of paths) $X_{t-} = \eta Y_t$. When $X$ jumps, it jumps down from $X_{t-} = \eta Y_t$ to

$$X_t = X_{t-} - Y_t(\eta - \xi) = \eta Y_t - Y_t(\eta - \xi) = \xi Y_t.$$

Observe also that $X_t + Y_t \geq (1 + \xi)Y_t > 0$ for $t \leq \tau^{\eta,\xi}$. The definition of the strategy (in terms of a gambling strategy $X_t$ and a stopping strategy $\tau$) is now as follows: $X \equiv X^{\eta,\xi}$ is given by (14) and $\tau = \tau^{\eta,\xi}$ by (15), at least on the set $\tau > 0$.

Note that since we are in the case $\gamma < 1$, $\max_{s \leq t} Y_s$ will be finite almost surely, and $A_\infty < \infty$. It is convenient to write $\overline{Y}_t = \max_{s \leq t} Y_s$. Let $H_z^Y$ denote the first hitting time of level $z$ by $Y$, and set $\tilde{A}_z = A_{H_z^Y}$. Recall that the situation has been reduced to the case $y \leq X_0/\eta$. Then, assuming $\overline{Y}_\infty \geq X_0/\eta$, for $X_0/\eta \leq z \leq \overline{Y}_\infty$ we have

$$\tilde{A}_z = \frac{\eta}{\eta - \xi} \int_0^{H_z^Y} \frac{dL_t}{Y_t} = \frac{\eta}{\eta - \xi} \int_{H_{X_0/\eta}^Y}^{H_z^Y} \frac{d\overline{Y}_t}{\overline{Y}_t}$$
$$= \frac{\eta}{\eta - \xi} \int_{X_0/\eta}^z \frac{du}{u} = \frac{\eta}{\eta - \xi} \ln\left(\frac{\eta z}{X_0}\right).$$

Informally, the strategy $(\tau^{\eta,\xi}, X^{\eta,\xi})$ can be described as follows. If $X_t \leq \xi Y_t$, then we stop immediately. We call this the stopping region $\mathcal{S}$. If $\xi Y_t < X_t < \eta Y_t$ (and especially if $\xi y < x < \eta y$), then we take a fair gamble such that $X_t$ jumps to an end of this interval immediately. We call this the gambling region $\mathcal{G}$. While $X_t > \eta Y_t$, we set $X_t$ to be constant. We call this the waiting region $\mathcal{W}$. If the bivariate process $(X, Y)$ is in $\mathcal{W}$, then either $Y$ eventually reaches the level $X_t/\eta$ and we reach the boundary between the gambling and waiting regions, or not, in which case $Y$ tends to zero (provided $\gamma < 1$), and $\tau = \infty$. If the former, then on the boundary between $\mathcal{G}$ and $\mathcal{W}$ we take a gamble which either makes the bivariate process jump to the boundary between $\mathcal{S}$ and $\mathcal{G}$ (and then we stop) or pushes the process (an infinitesimally small distance) back into the waiting region $\mathcal{W}$. (It is at this point that the argument is informal; the precise description is given via local times as above.) In this way, after perhaps an initial jump at time zero, once the process is in (the closure of) $\mathcal{W}$, it stays in this region, either indefinitely, or until there is a jump from the boundary of $\mathcal{W}$ to $\mathcal{S}$ at which point we stop. In particular, at no point do we ever enter the interior of $\mathcal{G}$. See Figure 1.

Now that we have described the martingale $(X_t)_{t \leq \tau}$, the next step is to determine the value function associated with this strategy.

PROPOSITION 7.  *Suppose $\gamma < R + \eta/(\eta - \xi)$. Under the above strategy, specified by the thresholds $(\xi, \eta)$, the value function is given by*

$$V^{\eta,\xi}(x,y) = \begin{cases} U_R(x+y), & x \leq \xi y, \\ J(x,y), & \xi y < x < \eta y, \\ K(x,y), & \eta y \leq x, \end{cases}$$

*where*

$$J(x,y) = \left(\frac{x}{\xi y} - 1\right)\frac{\xi}{\eta - \xi}K(\eta y, y)$$

(16)

$$+ \left(1 - \frac{x}{\eta y}\right)\frac{\eta}{\eta - \xi}\left[\frac{y^{1-R}(1+\xi)^{1-R} - 1}{1 - R}\right]$$

*and*

(17) $$K(x,y) = \frac{x^{1-R} - 1}{1 - R} + y^{1-\gamma}x^{\gamma - R}\Theta$$

*with*

$$\Theta \equiv \Theta_R(\eta, \xi) = \eta^{1-\gamma}\left[\frac{\eta - \xi + \eta(\eta^{R-1}(1+\xi)^{1-R} - 1)/(1 - R)}{\eta + (\eta - \xi)(R - \gamma)}\right].$$

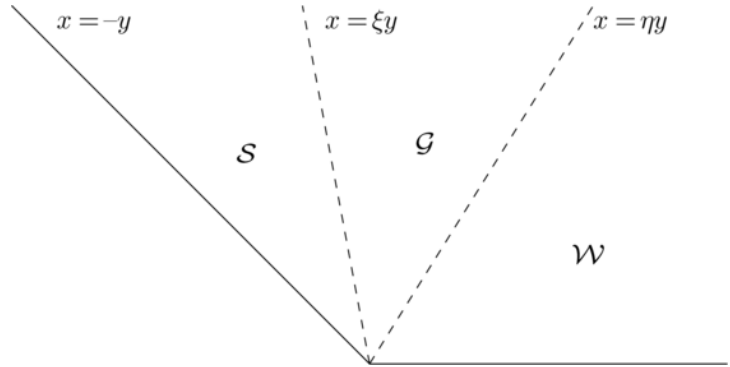

FIG. 1.  *Representation of the candidate strategy. In $\mathcal{S}$ we stop immediately, in $\mathcal{W}$ $X_t$ is constant and we wait until $Y_t = \eta X_t$ if ever, and in $\mathcal{G}$ we gamble, so that $X$ leaves the interval $(\xi Y_t, \eta Y_t)$ instantly.*

Now suppose $\gamma \geq R + \eta/(\eta - \xi)$. Then $V^{\eta,\xi}(x,y) = U_R(x+y)$ for $x \leq \xi y$ and infinity otherwise.

PROOF. The fundamental problem is to calculate
$$V^{\eta,\xi}(\eta y, y) = \mathbb{E}^{\eta y, y}[U_R(X_\tau + Y_\tau)]$$
where $\tau = \tau^{\eta,\xi}$ is the stopping rule defined above and $X = X^{\eta,\xi}$. The values for other starting positions can then be calculated easily using the martingale property in the gambling region, and the probability of reaching a new maximum in the waiting region. In particular, for $\xi y < x < \eta y$ we have
$$V^{\eta,\xi}(x,y) = \left(\frac{x}{\xi y} - 1\right)\frac{\xi}{\eta - \xi}V^{\eta,\xi}(\eta y, y) + \left(1 - \frac{x}{\eta y}\right)\frac{\eta}{\eta - \xi}U_R(\xi y, y),$$
and for $\eta y < x$ we have
$$V^{\eta,\xi}(x,y) = \left(\frac{\eta y}{x}\right)^{1-\gamma} V^{\eta,\xi}(x, x/\eta) + \left[1 - \left(\frac{\eta y}{x}\right)^{1-\gamma}\right]\frac{x^{1-R} - 1}{1 - R}.$$

So, suppose $(X_0, Y_0) = (\eta y, y)$. Recalling the definition of $\tilde{A}_z$, and using $\gamma < 1$, we have
$$\mathbb{P}(\overline{Y}_\tau > z) = \mathbb{P}(N_{\tilde{A}_z} = 0 | \overline{Y}_\infty > z)\mathbb{P}(\overline{Y}_\infty > z)$$
$$= \exp(-\tilde{A}_z)\left(\frac{y}{z}\right)^{1-\gamma} = \left(\frac{y}{z}\right)^\phi$$
where $\phi = (1 - \gamma) + (\eta/(\eta - \xi))$. Further,
$$\mathbb{P}(\overline{Y}_\tau \in dz, \tau = \infty) = \exp(-\tilde{A}_z)\mathbb{P}(\overline{Y}_\infty \in dz)$$
$$= (1 - \gamma)\left(\frac{y}{z}\right)^\phi \frac{dz}{z}$$
and on this event $X_\tau = \eta z$ and $Y_\tau = 0$. It follows that
$$\mathbb{P}(\overline{Y}_\tau \in dz, \tau < \infty) = \frac{\eta}{\eta - \xi}\left(\frac{y}{z}\right)^\phi \frac{dz}{z}$$
and on this set $X_\tau = \xi z$ and $Y_\tau = z$.

If we assume that $\gamma < R + \eta/(\eta - \xi)$, then $R + \phi > 1$ and
$V^{\eta,\xi}(\eta y, y)$
$$= \mathbb{E}^{\eta y, y}[U_R(X_\tau + Y_\tau)]$$
$$= \int_y^\infty \left(\frac{y}{z}\right)^\phi \frac{dz}{z}\left[(1-\gamma)\frac{(\eta z)^{1-R} - 1}{1 - R} + \frac{\eta}{\eta - \xi}\frac{(\xi z + z)^{1-R} - 1}{1 - R}\right]$$
$$= y^\phi \int_y^\infty \frac{dz}{(1 - R)}\left[\left((1-\gamma)\eta^{1-R} + \frac{\eta}{\eta - \xi}(1+\xi)^{1-R}\right)z^{-R-\phi} - \phi z^{-(1+\phi)}\right]$$
$$= \frac{y^{1-R}\eta^{1-R}(1 + (1-R)\eta^{-(1-\gamma)}\Theta) - 1}{1 - R},$$

where the last equality relies on a small amount of algebra.

If $\gamma \geq R + \eta/(\eta - \xi)$, then the integral near infinity of $z^{-R-\phi}$ is infinite.
□

4.2. *Optimal parameter choice.* It is clear from the expression for $K$ that in order to maximize the value function it is necessary to choose $\eta$ and $\xi$ to maximize $\Theta = \Theta_R(\eta, \xi)$. We maximize $\Theta$ over the set $\eta \geq 0$ and $-1 < \xi \leq \eta$. The case $\eta = \xi$ corresponds to the choice $\tau = \inf\{u : Y_u \leq x/\eta\}$ which was the class of stopping times considered in Section 3.

Recall the definition of $\gamma_-(R)$ from Definition 1. In this section we will generally assume that $\gamma < R$ so that the condition $\gamma < R + \eta/(\eta - \xi)$ is automatically satisfied.

LEMMA 8. *Suppose $R < 1$. For $0 < \gamma \leq \gamma_-(R)$ the maximum of $\Theta_R(\eta, \xi)$ is attained at $\xi^*(\gamma, R) = \eta^*(\gamma, R) = w^*$ where $w^*$ is the solution to (9).*

*For $\gamma_-(R) < \gamma < R$ the maximum of $\Theta_R(\eta, \xi)$ over $\xi > -1$, $\eta > 0$ and $\eta \geq \xi$ is attained at $(\eta^*, \xi^*)$ where $\eta^*$ and $\xi^*$ are given by*

$$\text{(18)} \quad \eta^* = \frac{(R-\gamma)(1-R)}{R}\left[(R+1-\gamma) - (R-\gamma)\left(\frac{1+R-\gamma}{1-\gamma}\right)^{1/R}\right]^{-1}$$

*and*

$$\text{(19)} \quad \xi^* = \frac{(R+1-\gamma)}{(R-\gamma)}\eta^* - \frac{1}{R}.$$

*In this case,*

$$\Theta_R(\eta^*, \xi^*) = \frac{(\eta^*)^{(1-\gamma)}R}{(R-\gamma)(1+R-\gamma)}.$$

PROOF. For fixed $\gamma \in (0, R)$ we want to maximize $\Theta_R(\eta, \xi)$. It is convenient to reparameterize the independent variables as $\eta$ and $\delta = (\eta - \xi)/\eta$, so that $0 < \delta < 1 + 1/\eta$. Then, if $\Phi(\eta, \delta) = \Theta_R(\eta, \xi)$, we have

$$\Phi(\eta, \delta) = \frac{\eta^{1-\gamma}}{1 + \delta(R-\gamma)}\left[\delta + \frac{(1-\delta+1/\eta)^{1-R} - 1}{1-R}\right].$$

For fixed $\delta$ it is clear that the maximum of $\Phi$ over $\eta$ is attained at an interior point. However, this need not be the case for fixed $\eta$, and the maximum value of $\Phi$ may occur at $\delta = 0$. Thus we have to investigate the possibility of maxima of $\Phi$ which occur on the boundaries. We have

$$\text{(20)} \quad \begin{aligned}\frac{\partial \Phi}{\partial \eta} &= \frac{\eta^{-\gamma}}{1 + \delta(R-\gamma)} \\ &\quad \times \left[(1-\gamma)\left(\delta + \frac{(1-\delta+1/\eta)^{1-R} - 1}{1-R}\right) - \frac{1}{\eta(1-\delta+1/\eta)^R}\right]\end{aligned}$$

and

$$\frac{\partial \Phi}{\partial \delta} = \frac{\eta^{1-\gamma}}{(1+\delta(R-\gamma))^2(1-R)}$$
$$\times \left[ (1-\gamma) - \left( \frac{(1-\gamma) - \delta R(R-\gamma) + (R-\gamma)/\eta}{(1-\delta+1/\eta)^R} \right) \right]. \tag{21}$$

Setting both expressions equal to zero, and obtaining expressions for $(1-\delta+1/\eta)^R$ in each case, we find that at a turning point

$$(1-R)\delta\{\delta\eta R(R-\gamma) - (R-\gamma-\eta R)\} = 0.$$

Hence, for fixed $\eta$, there are at most two turning points given by

$$\delta_1 = 0 \quad \text{and} \quad \delta_2 = \frac{R-\gamma-\eta R}{\eta(R-\gamma)R}.$$

Note that $0 < \delta_2$ only if $\eta < (R-\gamma)/R$.

Consider the turning point corresponding to $\delta_1 = 0$. For $\delta = 0$,

$$\frac{\partial \Phi}{\partial \eta} = \eta^{-\gamma} \left[ (1-\gamma) \frac{(1+1/\eta)^{1-R} - 1}{1-R} - \frac{1}{\eta(1+1/\eta)^R} \right].$$

Hence [recall (9)], $(\eta = w^*(\gamma), \delta = 0)$ is a turning point of $\Phi$. In order to determine whether this point is a (local) maximum it is necessary to consider the Hessian matrix of second derivatives. This is given by

$$\Phi''|_{(w^*,0)} = \frac{(w^*)^{-(1+\gamma)}}{(w^*+1)(w^* + (R-\gamma)/(1-\gamma))}$$
$$\times \begin{bmatrix} -\{R - \gamma(w^*+1)\} & \{R - \gamma(w^*+1)\}w^* \\ \{R - \gamma(w^*+1)\}w^* & -R(w^*)^4 \end{bmatrix}.$$

This matrix is negative definite, and hence the turning point is a local maximum, provided $0 < R - \gamma(w^*+1) < R(w^*)^2$.

The first inequality follows from Lemma 5. For the second, note that for small $\gamma$, $w^*(\gamma)$ is greater than 1 and $R(w^*)^2 > R - \gamma(w^*+1)$. This will remain the case until we first have $R(w^*)^2 = R - \gamma(w^*+1)$ or equivalently $w^* = (R-\gamma)/R$. Using the definition of $\Gamma_-$ the necessity condition for a local maximum translates to $\gamma < \gamma_-$. Otherwise, for $\gamma > \gamma_-$, $(w^*(\gamma), 0)$ is a saddle point.

Now consider the value $\delta = (R-\gamma-\eta R)/(\eta(R-\gamma)R)$. Substituting this expression into $\partial \Phi/\partial \delta = 0$ we find

$$\left( \frac{1+R-\gamma}{R-\gamma} - \frac{1-R}{R\eta} \right)^R = \frac{(1+R-\gamma)}{(1-\gamma)}.$$

We denote the solution of this equation by $\eta^*$ where $\eta^*$ is given in (18). For $\delta > 0$, or equivalently $\eta < (R-\gamma)/R$, we see that we need $\gamma > \gamma_-$.

Using $\xi = \eta(1 - \delta)$ and $\delta = (R - \gamma - \eta R)/(\eta(R - \gamma)R)$ we obtain (19). Finally, substitution of $\eta^*$ and $\xi^*$ into $\Theta_R(\eta, \xi)$ gives that $\Theta_R(\eta^*, \xi^*) = (\eta^*)^{1-\gamma} R/((R - \gamma)(1 + R - \gamma))$.

Note that for all values of $\gamma$ with $0 < \gamma < R$ we have shown that there is at most one local maximum. It follows from consideration of $\Phi$ on the boundaries that this is indeed the global maximum. $\square$

4.3. *The value function.* Now we can state the solution to the optimal stopping problem (4). For $\gamma \leq 0$ the optimal stopping rule is to stop immediately, and for $\gamma > R$, there is no optimal stopping rule (in the sense any candidate stopping rule which is finite can be improved upon by waiting longer). These results are exactly as in the case of Section 3, and so attention switches to the case $0 < \gamma < R$. The content of the next proposition is that for $\gamma$ in this range the optimal stopping rule is of the form described before Proposition 7, where the values of $\xi$ and $\eta$ are chosen to maximize $\Theta(\eta, \xi)$.

PROPOSITION 9. *Suppose $R < 1$.*

(i) *For $\gamma \leq 0$, $V_*^g(x, y) = U_R(x + y)$.*
(ii) *For $0 < \gamma \leq \gamma_-$ the value function is given by $V_*^g(x, y) = V^{w^*}(x, y)$ where $w^*$ solves* (9).
(iii) *For $\gamma_- < \gamma < R$ the value function is given by $V_*^g(x, y) = V^{\eta^*, \xi^*}(x, y)$ where $\eta^*$ and $\xi^*$ are given by (18) and (19).*
(iv) *For $\gamma \geq R$ the value function is given by $V_*^g(x, y) = \infty$.*

PROOF. The proof in the cases (i) and (ii) follows the proof of Proposition 6, the only additional step being to check that the value function is concave in $x$.

In case (iv) we know from Section 3 that if $x \geq 0$ and $\gamma > R$, then there is a sequence of strategies for which wealth is constant and for which the value function is arbitrarily large. Since this strategy remains feasible in the case with gambling opportunities, the value function must be infinite in this case also. If $x < 0$, but $x + y > 0$, then it is possible to take an initial gamble such that with some probability the post-gamble wealth $X_0$ is positive; and conditional on this event the value function is infinite. In this way we conclude that $V^g(x, y) = \infty$ for $x > -y$ also. Finally, suppose $\gamma = R$. Let $\eta = \varepsilon$ and $\xi = -1 + \varepsilon$. Then $\Theta_R(\varepsilon, -1 + \varepsilon) = \varepsilon^{-\gamma}$ and for $x > 0$, $V^g(x, y) \geq V^{\varepsilon, -1+\varepsilon}(x, y) > y^{1-\gamma} x^{\gamma - R} \Theta_R(\varepsilon, -1 + \varepsilon) - 1/(1 - R)$. Hence, for $x > 0$, $V^g(x, y) = \infty$ and this can be extended to $x > -y$ by considering a suitable initial gamble as before.

It remains to prove the result in the case $\gamma_-(R) < \gamma < R$. Proposition 7 gives the value function for candidate strategies of a given form. The results of Lemma 8 describe how to choose the optimal member of this class. It remains to show that it is sufficient to restrict attention to strategies of this type.

Write $V^c$ as shorthand for $V^{\eta^*,\xi^*}$ where the superscript is intended to denote the fact that this is a conjectured value function. Then for $x \leq \xi^* y$, $V^c(x,y) = U_R(x+y)$; for $\xi^* y < x \leq \eta^* y$,

$$V^c(x,y) = \frac{1}{1-R}\left[\left(\frac{y^{1-R}}{\eta^* - \xi^*}\left\{\left(\frac{x}{y} - \xi^*\right)(\eta^*)^{1-R}\frac{(2R-\gamma)(1-\gamma)}{(R-\gamma)(1+R-\gamma)}\right.\right.\right.$$
$$\left.\left.\left. + \left(\eta^* - \frac{x}{y}\right)(1+\xi^*)^{1-R}\right\}\right) - 1\right]$$

and for $x \geq y\eta^*$,

$$V^c(x,y) = \frac{x^{1-R} - 1}{1-R} + y^{1-\gamma}x^{-(R-\gamma)}(\eta^*)^{1-\gamma}\frac{R}{(R-\gamma)(1+R-\gamma)}.$$

Since $V^c$ is the value function associated with some admissible strategy [indeed the strategy associated with $X$ and $\tau$ as defined in (14) and (15) for the constants $\eta^*$ and $\xi^*$], $V^c$ is a lower bound on the value function $V_*^g$.

In order to show that $V^c \equiv V_*^g$ it will be sufficient to show that $V^c(x,y) \geq U_R(x+y)$ and that $V^c(X_t, Y_t)$ is a supermartingale for $t \leq \tau$, where $\tau$ is any stopping rule such that $(X_t + Y_t) > 0$ for all $t < \tau$, and $X \in \mathcal{X}$. This rules out Ponzi schemes for the wealth process $X_t$.

The verification that $V^c(x,y) \geq U_R(x+y)$ is a lengthy but straightforward exercise. To prove that $V^c(X_t, Y_t)$ is a supermartingale for $t \leq \tau$ we need the following lemma.

LEMMA 10. *Suppose $\gamma_-(R) < \gamma < R$, and suppose $V^c = V^{\eta^*,\xi^*}$. Then $V^c$ is concave in $x$ so*

(22) $$V^c(x+\Delta, y) - V^c(x,y) \leq \Delta V_x^c(x,y),$$

*and $L^Y V^c \leq 0$ where*

(23) $$L^Y V = \mu y V_y + \tfrac{1}{2}\sigma^2 y^2 V_{yy}.$$

*Moreover, the first derivatives with respect to $x$ and $y$ of $V^c$ are continuous at $x = \xi^* y$ and $x = \eta^* y$.*

PROOF. Define the selling, gambling and waiting regions $\mathcal{S}$, $\mathcal{G}$ and $\mathcal{W}$ via $\mathcal{S} = \{(x,y): y > 0, -y < x \leq \xi^* y\}$, $\mathcal{G} = \{(x,y): y > 0, \xi^* y < x < \eta^* y\}$ and $\mathcal{W} = \{(x,y): y > 0, \eta^* y \leq x\}$ as above.

*Concavity in $x$*: In $\mathcal{S}$ we have $V_x^c = (x+y)^{-R}$ and $V_{xx}^c = -R(x+y)^{-(R+1)} < 0$. In $\mathcal{G}$, $V^c$ is linear in $x$ with derivative $V_x^c = y^{-R}((\eta^*)^{1-R} + (\eta^*)^{1-\gamma}\Theta^* - (1+$ $\xi)^{1-R})/((1-R)(\eta^*-\xi^*))$. Finally in $\mathcal{W}$, $V_x^c = x^{-R} - (R-\gamma)x^{-(1+R-\gamma)}y^{1-\gamma}\Theta^*$ and $V_{xx}^c = -Rx^{-(R+1)}(1-(y\eta^*/x)^{1-\gamma}) \leq 0$.

$L^Y V^c \leq 0$: In $\mathcal{S}$ we have $L^Y V^c = (x+y)^{-(R+1)}\gamma y^2((x/y) - (R-\gamma)/\gamma)$. This is negative provided $\xi^* \leq (R-\gamma)/\gamma$ which follows from Lemma 5 and the fact that $\xi^* \leq w^*$. In $\mathcal{G}$, $L^Y V^c = y^{1-R}\eta^{-R}(1-\gamma)R((x/y) - \eta^*) < 0$. Finally, in $\mathcal{W}$, $L^Y V^c = 0$.

*Continuity of the derivatives at the boundaries $\mathcal{S}/\mathcal{G}$ and $\mathcal{G}/\mathcal{W}$.* Continuity of the derivatives $V_x^c$ and $V_y^c$ follows from the identities

$$\frac{(1-\gamma)}{(1+R-\gamma)}(\eta^*)^{1-R} = (1+\xi^*)^{-R}$$

and

$$(1+\xi^*)^{-R} = \frac{1}{(1-R)(\eta^*-\xi^*)}\left[(\eta^*)^{1-R}\left(\frac{(2R-\gamma)(1-\gamma)}{(R-\gamma)(1+R-\gamma)}\right) - (1+\xi^*)^{1-R}\right]. \quad \square$$

Return to the proof of Proposition 9. Applying Itô's formula to $V^c$ (note that we need a version of Itô's formula which applies to functions of discontinuous martingales; see, e.g., Rogers and Williams [11], Theorem VI.39.1) and using (22) and then (23), for any $X \in \mathcal{X}$,

$$V^c(X_{t\wedge\tau}, Y_{t\wedge\tau}) \leq V^c(x,y) + \int_0^{t\wedge\tau} \sigma Y_u V_y^c \, dW_u$$
$$+ \int_0^{t\wedge\tau} (\mu Y_u + \tfrac{1}{2}\sigma^2 Y_u^2 V_{yy}^c) \, du + \int_0^{t\wedge\tau} V_x^c(X_{u-}, Y_u) \, dX_u$$
(24)
$$\leq V^c(x,y) + \int_0^{t\wedge\tau} \sigma Y_u V_y^c \, dW_u + \int_0^{t\wedge\tau} V_x^c(X_{u-}, Y_u) \, dX_u$$
$$= V^c(x,y) + M_{t\wedge\tau}$$

where $M$ is a local martingale.

Since $R < 1$ it follows easily that $V^c$ is bounded below on the domain $x+y \geq 0$. Hence the local martingale $M_{t\wedge\tau}$ is a supermartingale and this property is inherited by $V^c(X_{t\wedge\tau}, Y_{t\wedge\tau})$. In particular $V^c(X_{t\wedge\tau}, Y_{t\wedge\tau})$ converges almost surely, and so does $X_{t\wedge\tau}$. Then by Fatou,

$$\mathbb{E}[U_R(X_\tau + Y_\tau)] \leq \liminf \mathbb{E}[V^c(X_{t\wedge\tau}, Y_{t\wedge\tau})] \leq V^c(x,y)$$

and it follows that $V^c \equiv V_*^g$. $\quad\square$

Note that Theorem 2 follows immediately on comparison of Proposition 6 with Proposition 9.

**5. Coefficients of relative risk aversion larger than 1.** In this section we consider the problems (4) and (6) for the case $R > 1$. The case $R > 1$ introduces two new elements into the analysis.

The first new element is the fact that the case $\gamma > R$ becomes more complicated. When $R < 1$ and $\gamma > R$ the value function is infinite. In this case the mean value of $Y^{1-R}$ is growing, and there is both no great risk associated with small (and even zero) values of $(X + Y)$ and a great reward from large values. However, when $R > 1$, the agent must avoid $X_t + Y_t = 0$ at all costs and the rewards for large $(X + Y)$ are bounded above. The net effect is that when wealth $x$ is fixed and negative, then it can never be optimal to allow $x + Y_t$ to hit zero, and there is a new nondegenerate solution in the case $\gamma > R > 1$. For example, for the problem in (6) we have:

THEOREM 11. *Suppose $\gamma > R > 1$.*
*Suppose $x \geq 0$. Then $V_*^n(x, y) = U(\infty) = 1/(R-1)$.*
*Suppose $x < 0$. Set $z^* = (\gamma - 1)/(\gamma - R) > 1$ and for $|x| < y < |x|z^*$ define*

$$V^{z^*}(x, y) = \frac{1 - (x+y)^{-(R-1)}}{R-1}$$

*and for $|x|z^* \leq y$ define*

$$V^{z^*}(x, y) = \frac{1}{R-1}\left[1 - \frac{(\gamma-1)^{\gamma-1}}{(R-1)^{R-1}(\gamma-R)^{\gamma-R}}|x|^{\gamma-R}y^{-(\gamma-1)}\right].$$

*Then $V_*^n(x, y) = V^{z^*}(x, y)$.*

Theorem 11 can be proved in a similar fashion to other results in this paper. However, we will not discuss the case $\gamma > R > 1$ in detail. Instead we will concentrate on the second effect of considering the case $R > 1$, which is that the fact that the objective function is not bounded below introduces new complications.

The definitions of the critical ratios and the critical strategy do not change when we consider $R > 1$, except that now we restrict attention to $\gamma < 1$ rather than $\gamma < R$. As we shall see, the critical value of $\gamma$ at which gambling becomes useful is still given by the solution to $\Gamma_R(\gamma) = 0$, and $\eta^*$, $\xi^*$, $\Theta^*$ and the value function are still given by their expressions in Lemma 8. Furthermore, the proofs of Proposition 7 and Lemma 8 are unchanged [except that $\gamma < R + \eta/(\eta - \xi)$ is now automatic] and the lower bound parts of the verification lemmas (Propositions 6 and 9) are also valid. The only changes are in the proofs of the upper bounds.

We concentrate on the upper bound in Proposition 9, since the situation in Proposition 6 is simpler and can be proved by identical ideas. The key issue is that when $R < 1$ we can easily conclude that the local martingale $M$ in (24) is bounded below, and hence a supermartingale. When $R > 1$ this is no longer the case.

There is an easy way to finesse the problem, which is to modify the definition of $\mathcal{X}$ so that admissible pairs $(\tau, X)$ must satisfy $X_t + Y_t > \varepsilon > 0$ or more generally $\mathbb{E}[(\inf_{t \leq \tau}\{X_t + Y_t\})^{1-R}] < \infty$. In that case $M_{t \wedge \tau}$ is bounded below by an integrable random variable and hence a supermartingale. (A further alternative to the same effect would be to require that $(\{X_{t \wedge \tau} + Y_{t \wedge \tau}\})^{1-R}$ is uniformly integrable; see [4], Lemma 5.2.) However, this modification is unsatisfactory since it arbitrarily rules out strategies which should in any case be suboptimal—the optimal strategy involves liquidation before $X_t + Y_t$ gets too small—and it rules these out artificially rather than by this suboptimality property.

Instead, we retain the definition of $\mathcal{X}$ so that $X_t + Y_t \geq 0$ for $t \leq \tau$. We want to find $V_*^g \equiv V_*^g(x_0, y_0)$ for particular initial values $X_0 = x_0$ and $Y_0 = y_0$. Typically we solve this problem by finding $V_*^g(x, y)$ for all initial values $(x, y)$ simultaneously. Now we have to be slightly more careful. Given initial values $(x_0, y_0)$, we replace the objective function $U$ with a larger function $\tilde{U}(x, y)$, which is bounded below. We now find the candidate value function $\tilde{V}$ associated with $\tilde{U}$ for all possible initial starting points. Since $\tilde{U}$ is bounded below, we can prove that the conjectured value function $\tilde{V}$ is the true value function $\tilde{V}_*$ for objective function $\tilde{U}$. Finally, since $\tilde{U}$ is chosen so that it agrees with $U$ on the stopping set, we conclude the result we want that $V^{\eta^*, \xi^*}(x_0, y_0) = V_*^g(x_0, y_0)$. The key fact that makes this approach work is that for the optimal strategy the pair $(X_t, Y_t)$ is always stopped before getting close to the origin, or to the line $x = -y$. For this reason we can change the value of $U$ on these neighborhoods without altering the value function at the starting point.

LEMMA 12. *Suppose $R > 1$ and $\gamma_-(R) < \gamma < 1$. Fix $\tilde{y} > 0$ and define $\tilde{\mathcal{D}} = \{(x, y) : 0 < y \leq \tilde{y}, -y \leq x \leq \eta \tilde{y}\}$. Define $\tilde{\mathcal{S}} = \mathcal{S} \cap \tilde{\mathcal{D}}^C$, $\tilde{\mathcal{G}} = \mathcal{G} \cap \tilde{\mathcal{D}}^C$ and $\tilde{\mathcal{W}} = \mathcal{W} \cap \tilde{\mathcal{D}}^C$, where $\tilde{\mathcal{D}}^C$ denotes the complement of $\tilde{\mathcal{D}}$.*

(i) *Define $\tilde{U}$ as follows: $\tilde{U} = U$ on $\tilde{\mathcal{G}}$ and $\tilde{\mathcal{W}}$; on $\tilde{\mathcal{S}}$,*

$$\tilde{U}(x, y) = \frac{1}{1-R}\left[\left(\frac{y^{1-R}}{\eta^* - \xi^*}\left\{\left(\frac{x}{y} - \xi^*\right)(\eta^*)^{1-R}\frac{(2R-\gamma)(1-\gamma)}{(R-\gamma)(1+R-\gamma)}\right.\right.$$
$$\left.\left. + \left(\eta^* - \frac{x}{y}\right)(1+\xi^*)^{1-R}\right\}\right) - 1\right]$$

*and on $\tilde{\mathcal{D}}$,*

$$\tilde{U}(x, y) = K^*(\tilde{y}\eta^*, y) - (\eta^*\tilde{y} - x)K_x^*(\tilde{y}\eta^*, y)$$
$$= \frac{R}{1-R}\tilde{y}^{1-R}(\eta^*)^{1-R} + x\tilde{y}^{-R}(\eta^*)^{-R}$$

$$+ \frac{R}{(1+R-\gamma)} y^{1-\gamma} \tilde{y}^{-(R-\gamma)} (\eta^*)^{1-R} \left( \frac{(1+R-\gamma)}{(R-\gamma)} - \frac{x}{\eta^* \tilde{y}} \right) - \frac{1}{1-R}$$

*where $K^*$ is the function in (17) evaluated at $\eta^*$ and $\xi^*$. Then $\tilde{U} \geq U$ and $\tilde{U}$ is bounded below.*

*(ii) Define $\tilde{V}$ via $\tilde{V} = \tilde{U}$ on $\tilde{\mathcal{S}}$ and $\tilde{\mathcal{D}}$ and $\tilde{V} = V^{\eta^*,\xi^*}$ on $\tilde{\mathcal{G}}$ and $\tilde{\mathcal{W}}$. Then $\tilde{V}(X_t, Y_t)$ is a supermartingale, and for all $X \in \mathcal{X}$,*

(25) $$\tilde{V}(x,y) \geq \sup_{\tau, X \in \mathcal{X}} \mathbb{E}^{x,y}[\tilde{U}(X_\tau, Y_\tau)].$$

PROOF. (i) On $\tilde{\mathcal{S}}$, $\tilde{U}$ is constructed such that the value and first $x$-derivative match those of $U$ at $\xi^* x$. Since $\tilde{U}$ is linear in $x$, whereas $U$ is concave, we have $\tilde{U} \geq U$ on $\mathcal{S}$.

Similarly, on $\tilde{\mathcal{D}}$, $\tilde{U}$ is constructed such that the value and first $x$-derivative match those of $V^c$ at $(\eta \tilde{y}, y)$. Since $\tilde{U}$ is linear in $x$, whereas $V^c$ is concave, we have $\tilde{U} \geq V^c \geq U$ on $\tilde{\mathcal{D}}$.

(ii) Exactly as in the proof of Proposition 9, it will follow that $\tilde{V}(X_t, Y_t)$ is a supermartingale provided that $\tilde{V}$ is convex in $x$, $L^Y \tilde{V} \leq 0$ and the first derivatives of $\tilde{V}$ are continuous at the boundaries between $\tilde{\mathcal{S}}$, $\tilde{\mathcal{G}}$, $\tilde{\mathcal{W}}$ and $\tilde{\mathcal{D}}$. It then follows that

$$\tilde{V}(X_t, Y_t) \leq \tilde{V}(x,y) + \tilde{M}_t$$

[recall (24)], where $\tilde{M}_t$ is a local martingale which is bounded below, and hence a supermartingale.

Let $J^*(x,y)$ and $K^*(x,y)$ denote the functions $J$ and $K$ defined in Proposition 7, but evaluated using the optimal parameters $\eta^*$ and $\xi^*$. Then in $\tilde{\mathcal{S}}$, $\tilde{U}(x,y) = J^*(x,y)$ where the domain of $J^*$ has been extended from $\tilde{\mathcal{G}}$ into $\tilde{\mathcal{S}}$. It follows that $\tilde{V}$ is linear in $\tilde{\mathcal{S}}$ and $\tilde{\mathcal{G}}$, and in $\tilde{\mathcal{D}}$ linearity follows by construction. In $\tilde{\mathcal{W}}$ convexity of $\tilde{V}$ is guaranteed by Lemma 10.

Similarly, the fact that $L^Y \tilde{V} \leq 0$ follows immediately in $\tilde{\mathcal{S}}$, $\tilde{\mathcal{G}}$ and $\tilde{\mathcal{W}}$, and in $\tilde{\mathcal{D}}$ it is easy to show that $L^Y \tilde{V} = 0$.

Finally it is necessary to check that the first derivatives of $\tilde{V}$ are continuous at the boundaries. For the $x$-derivative on the $\tilde{\mathcal{W}}/\tilde{\mathcal{D}}$ boundary, this follows by definition and the only nontrivial derivative to check is the $y$-derivative on the $\tilde{\mathcal{G}}/\tilde{\mathcal{D}}$ boundary, or equivalently on the $\tilde{\mathcal{S}}/\tilde{\mathcal{D}}$ boundary. The equality of the derivatives on either side of the boundary follows by direct calculation, or by using the fact that in $\tilde{\mathcal{G}}$ and $\tilde{\mathcal{S}}$

$$\tilde{V}(x,y) = J^*(\eta^* y, y) - (\eta^* y - x) J^*_x(\eta^* y, y)$$

whereas in $\tilde{\mathcal{D}}$, the same formula remains true but with $J^*$ replaced by $K^*$. Using this second method, the result follows from the fact that $K^*_{xx}(\eta^* y, y) = 0$.

Given that $\tilde{V}(X_t, Y_t)$ is a supermartingale it is a short step to prove (25). Note that on $\tilde{\mathcal{G}}$ and $\tilde{\mathcal{W}}$,

$$\tilde{V} = V^{\eta^*, \xi^*} \geq U = \tilde{U}$$

so that $\tilde{V} \geq \tilde{U}$ everywhere. Then

$$\tilde{V}(x,y) \geq \sup_{\tau, X \in \mathcal{X}} \mathbb{E}[\tilde{V}(X_\tau, Y_\tau)] \geq \sup_{\tau, X \in \mathcal{X}} \mathbb{E}[\tilde{U}(X_\tau, Y_\tau)]. \qquad \square$$

PROPOSITION 13. *Suppose $R > 1$ and $\gamma < 1$.*

(i) *For $\gamma \leq 0$, $V_*^g(x,y) = U_R(x+y)$.*
(ii) *For $0 < \gamma \leq \gamma_-$ the value function is given by $V_*^g(x,y) = V^{w^*}(x,y)$.*
(iii) *For $\gamma_- < \gamma < 1$ the value function is given by $V_*^g(x,y) = V^{\eta^*, \xi^*}(x,y)$.*

This leads immediately to the following analogue of Theorem 2.

THEOREM 14. *Suppose $R > 1$ and $\gamma < 1$. For $\gamma \leq \gamma_-(R)$ we have that $V_*^n \equiv V_*^g$. Conversely, for $\gamma_-(R) < \gamma < 1$ we have that $V_*^n < V_*^g$.*

PROOF OF PROPOSITION 13 IN THE CASE $\gamma_- < \gamma$. Suppose $(x,y)$ is such that $x \geq \xi^* y$. Fix $\tilde{y} \leq y$, and use this $\tilde{y}$ to define $\tilde{\mathcal{D}}, \tilde{\mathcal{S}}, \tilde{\mathcal{G}}$ and $\tilde{\mathcal{W}}$ as in the lemma above, together with $\tilde{U}$ and $\tilde{V}$. Note that either $(x,y) \in \tilde{\mathcal{G}}$ or $(x,y) \in \tilde{\mathcal{W}}$.

We know from consideration of the strategy $(\tau^{\eta^*, \xi^*}, X^{\eta^*, \xi^*})$, which we abbreviate here to $(\tau^*, X^*)$, that

$$V^{\eta^*, \xi^*}(x,y) = \mathbb{E}[U(X^*_{\tau^*}, Y_{\tau^*})] \leq \sup_{\tau, X \in \mathcal{X}} \mathbb{E}[U(X_\tau, Y_\tau)] = V_*^g(x,y).$$

Now let $(\tau, X)$ be any admissible strategy. Then, by a simple comparison and Lemma 12,

$$\mathbb{E}[U(X_\tau, Y_\tau)] \leq \mathbb{E}[\tilde{U}(X_\tau, Y_\tau)] \leq \tilde{V}(x,y).$$

Finally, since $(x,y) \in \tilde{\mathcal{G}} \cup \tilde{\mathcal{W}}$, in which region $\tilde{V} = V^{\eta^*, \xi^*}$ we have $\mathbb{E}[U(X_\tau, Y_\tau)] \leq V^{\eta^*, \xi^*}(x,y)$ and $V_*^g = V^{\eta^*, \xi^*}$.

Now suppose that $x < \xi^* y$. For fixed risk aversion $R$, the optimal ratio $\xi^*$ is decreasing in $\gamma$. Let $\hat{\gamma} > \gamma$ be such that $\xi^*(\hat{\gamma}) = x/y$, and let $\hat{Y}$ denote the solution to (2) for this parameter value. Then

$$\hat{Y}_t = Y_t e^{\sigma^2(\hat{\gamma} - \gamma)t/2} \geq Y_t.$$

Thus

$$(26) \qquad \sup_{\tau, X \in \mathcal{X}} \mathbb{E}[U(X_\tau, Y_\tau)] \leq \sup_{\tau, X \in \mathcal{X}} \mathbb{E}[U(X_\tau, \hat{Y}_\tau)] = \frac{(x+y)^{1-R} - 1}{1-R},$$

where this last equality follows from the fact that if $x = \xi^*(\hat{\gamma})y$, then it is optimal to stop immediately. Since there is equality in (26) for $\tau = 0$ we have $V_*^g(x,y) = V^{\eta^*, \xi^*}(x,y) = U(x,y)$ for $x < \xi^* y$. $\square$

## 6. Conclusions and further remarks.

6.1. *Logarithmic utility.* The results for the case of logarithmic utility can easily be recovered by taking the limit $R \to 1$. For example, the equation $\Gamma_R(\gamma_-) = 0$ can be rewritten as

$$\frac{1}{R-1}\left[\left(\frac{(2R-\gamma_-)}{(R-\gamma_-)}\right)^{R-1} - 1\right] = \frac{R}{(1-\gamma_-)(2R-\gamma_-)}.$$

Letting $R \to 1$ we find that $\gamma_- = \gamma_-(1)$ is the unique solution in $(0,1)$ to

$$\ln\left(\frac{2-\gamma}{1-\gamma}\right) = \frac{1}{(1-\gamma)(2-\gamma)}.$$

In a similar fashion we can let $R \to 1$ in the defining equations for $w^*$, $V^w$, $\eta^*$, $\xi^*$ and $V^{\eta,\xi}$. In this way we can deduce the results for logarithmic utility from Propositions 9 and 13.

COROLLARY 15. *Suppose $U(x) = \ln x$ and $\gamma < 1$. For $\gamma \leq \gamma_-(1)$ we have that $V_*^n \equiv V_*^g$. Conversely, for $\gamma_-(1) < \gamma < 1$ we have that $V_*^n < V_*^g$.*

6.2. *Other limiting cases.* It is interesting to consider the limiting cases $R \downarrow 0$ and $R \uparrow \infty$. As $R \downarrow 0$ the utility function approaches linear and $\gamma_-(R)$ approaches zero. If $\gamma \leq 0$, then it is always optimal to sell the real asset immediately, whereas if $\gamma > 0$, it is optimal to hold onto the real asset indefinitely.

Conversely, in the limit $R \uparrow \infty$, $\gamma_-(R) \to 1$. In this case, for $\gamma < 1$ it is never optimal to gamble. Let $E(w) = \lim_{R \uparrow \infty}(R-1)U_R(1 + w/(R-1))$; then $E(w) = 1 - e^{-w}$. In this sense at least the limit $R \uparrow \infty$ corresponds to exponential utility. The optimal sale problem for exponential utility has been studied by Henderson [3]. For exponential utility wealth factors out of the problem, so the value function is always concave in $x$.

6.3. *Convexities and gambling in related models.* The main phenomenon which our model attempts to capture is that the timing option (discretionary stopping) and market incompleteness potentially induce a convexity in the value function, and this encourages the risk-averse agent to gamble.

Convexities of this form, and the consequent predictions of risk-seeking behavior, can arise in other ways. First, the objective function may itself be convex, for example if the agent has limited liability with respect to losses. (The agent then "gambles for resurrection.") Second, a convexity may arise from an interaction between discounting and consumption, as the following example, slightly modified from Karatzas and Wang [8], illustrates.

Consider the problem of finding, for a positive "discount factor" $\rho > 0$,

$$\text{(27)} \quad V_*^g \equiv V_*^g(x) = \sup_{\tau, X \in \mathcal{X}} \mathbb{E}[e^{-\rho\tau} \ln(X_\tau)],$$

where

$$\text{(28)} \quad \mathcal{X} = \{X_t : \mathbb{E}[X_0] = x; X_u \geq 0, \ u \leq \tau; X \text{ is a càdlàg martingale}\}.$$

Define also

$$V_*^n \equiv V_*^n(x) = \sup_\tau \mathbb{E}[e^{-\rho\tau} \ln(x)].$$

It is easy to see that

$$V_*^n(x) = (\ln x)^+.$$

(When $x < 1$ the agent can defer $\tau$ indefinitely, and the presence of the discount factor encourages him to do so.)

The function $V_*^n(x)$ is not convex. It follows that the agent who can undertake a gamble at time 0 should do so, and that

$$V_*^g(x) = (\ln x) I_{\{x > e\}} + (x/e) I_{\{x \leq e\}}.$$

For $x \geq e$, the optimal stopping rule is $\tau = 0$. For $x < e$ there is no optimal strategy, but there is a sequence of strategies indexed by $m \in \mathbb{N}$, with associated value functions which converge to $V_*^g$. These strategies involve an initial fair gamble at time 0, after which wealth is either $e$ or $x/m$ [with probabilities $p = x(m-1)/(em-x)$ and $1-p$, resp.].

It should be noted that the objective function $e^{-\rho\tau} \ln x$ has some perverse features, and the interpretation of $\rho$ as a discount factor is hard to justify in economic terms. When there is no uncertainty and $x > 1$, the agent prefers $\tau = 0$, since endowments received later are less valuable. However, when $x < 1$, the agent prefers to take $\tau$ as large as possible. In this case the agent prefers a (certain) payout later rather than sooner, which is inconsistent with the standard interpretation of discounting. Effectively, the fact that losses can be deferred indefinitely introduces a gambling for resurrection element into the problem. Note that since there is no discount factor in our problem, there is no relationship between the causes of the incentives to gamble in the Karatzas and Wang [8] model, and in the model of this paper.

The above examples illustrate that there are many reasons why agents facing optimization problems may seek to take outside gambles. Sometimes these reasons may be traced back to a convexity in the objective function. In contrast, in our model the incentive to gamble arises from an interaction between the timing option over when to sell, and the market incompleteness.

6.4. *Concluding remarks.* In this article we have given an explicit solution to a mixed optimal control/optimal stopping problem. Associated with the explicit solution for the value function is an explicit element $X \in \mathcal{X}$. This control can be characterized in terms of a local time on a ray. The optimal pair $(X_t, Y_t)$ receives a local time push to keep the process within the region $X_t \geq \eta^* Y_t$, and to preserve the martingale property of $X$ there are compensating downward jumps. Thus, even though the setup involves continuous processes, it is necessary to introduce discontinuous processes in order to define the optimal wealth process.

In the case of relative risk aversion coefficients less than 1, the proof can be completed by conjecturing the form of the optimal strategy, and then using a verification lemma. When $R > 1$ the fact that the objective function is unbounded below introduces significant extra complications. However, it is intuitively clear that the conjectured strategy is still optimal, since the optimal strategy is to sell immediately when wealth is small. This gives us the key to proving optimality in this case: we modify the objective function on parts of the space that the optimal controlled process never reaches; for this modified problem the value function is bounded everywhere, but the solution is equal to the solution of the original problem at the starting point.

The most interesting feature about the problem is that the solution is not as might be predicted. Instead, although the agent is fully rational and risk averse, the incompleteness of the market and the presence of the American-style timing option to sell the real asset induces her to gamble.

## REFERENCES

[1] DIXIT, A. K. and PINDYCK, R. S. (1994). *Investment under Uncertainty.* Princeton Univ. Press.
[2] EVANS, J. D., HENDERSON, V. and HOBSON, D. (2007). Optimal timing for an asset sale in an incomplete market. *Math. Finance.* To appear.
[3] HENDERSON, V. (2007). Valuing the option to invest in an incomplete market. *Math. Financ. Econ.* **1** 103–128. MR2365685
[4] JACKA, S. D. (1991). Optimal stopping and best constants for Doob-like inequalities. I. The case $p = 1$. *Ann. Probab.* **19** 1798–1821. MR1127729
[5] KARATZAS, I. and KOU, S. G. (1998). Hedging American contingent claims with constrained portfolios. *Finance Stoch.* **2** 215–258. MR1809521
[6] KARATZAS, I. and OCONE, D. (2002). A leavable bounded-velocity stochastic control problem. *Stochastic Process. Appl.* **99** 31–51. MR1894250
[7] KARATZAS, I. and SUDDERTH, W. D. (1999). Control and stopping of a diffusion process on an interval. *Ann. Appl. Probab.* **9** 188–196. MR1682584
[8] KARATZAS, I. and WANG, H. (2000). Utility maximization with discretionary stopping. *SIAM J. Control Optim.* **39** 306–329 (electronic). MR1780921
[9] MERTON, R. C. (1969). Lifetime portfolio selection under uncertainty: The continuous-time case. *Rev. Econom. Statist.* **51** 247–257.
[10] MÜLLER, A. and STOYAN, D. (2002). *Comparison Methods for Stochastic Models and Risks.* Wiley, Chichester. MR1889865

[11] ROGERS, L. C. G. and WILLIAMS, D. (2000). *Diffusions, Markov Processes, and Martingales. Vol. 2. Itô Calculus.* Reprint of the second (1994) edition. Cambridge Univ. Press, Cambridge. MR1780932

[12] SAMUELSON, P. A. (1965). Rational theory of warrant pricing. With an appendix by H. P. McKean, A free boundary problem for the heat equation arising from a problem in mathematical economics. *Industrial Management Review* **6** 13–31 and 32–39.

WARWICK BUSINESS SCHOOL  
UNIVERSITY OF WARWICK  
COVENTRY CV4 7AL  
UNITED KINGDOM  
E-MAIL: Vicky.Henderson@wbs.ac.uk

DEPARTMENT OF STATISTICS  
UNIVERSITY OF WARWICK  
COVENTRY CV4 7AL  
UNITED KINGDOM  
E-MAIL: d.hobson@warwick.ac.uk